\documentstyle[12pt,epsfig]{article}
\begin{document}

\newcommand{\al}{\mbox{$\alpha$}}
\newcommand{\be}{\mbox{$\beta$}}
\newcommand{\g}{\mbox{$\gamma$}}
\newcommand{\de}{\mbox{$\delta$}}
\newcommand{\om}{\mbox{$\omega$}}
\newcommand{\Om}{\mbox{$\Omega$}}
\newcommand{\tri}{\mbox{$\triangle$}}
\newcommand{\tr}{\mbox{$\tilde{r}$}}
\newcommand{\tom}{\mbox{$\tilde{\omega }$}}

\def\beq{\begin{equation}}
\def\eeq{\end{equation}}
\def\beqa{\begin{eqnarray}}
\def\eeqa{\end{eqnarray}}
\def\s{\sigma}
\def\ve{\varepsilon}
\def\bl{\bar{\lambda}}
\def\da{\dagger}
\def\la{\lambda}
\def\ti{\tilde}
\def\pr{\prime}
\def\f{\frac}
\def\sq{\sqrt}
\def\pa{\partial}
\def\so{{\scriptsize \Omega}}
\def\somh{{\scriptsize \Omega_H}}
\def\+{{\scriptsize \Omega_{+}}}
\def\-{{\scriptsize \Omega_{-}}}
\def\0{{\scriptsize \Omega_{0}}}
\def\b{{\scriptsize \beta}}

\begin{titlepage}
\begin{center}
   \vskip 5em
   {\LARGE Superradiance and the Statistical-Mechanical Entropy of Rotating
BTZ Black Holes}
   \vskip 1.5em
   {\large Jeongwon Ho\footnote{jwho@physics3.sogang.ac.kr} and
           Gungwon Kang\footnote{kang@theory.yonsei.ac.kr}
   \\[.5em]}
{\em Department of Physics and Basic Science Research Institute,
Sogang University, C.P.O. Box 1142, Seoul 100-611, Korea
}  \\[.7em]
\end{center}
\vskip 1em

\begin{abstract}

We have considered the divergence structure in the brick-wall model
for the statistical mechanical entropy of a quantum field in thermal
equilibrium with a black hole which {\it rotates}. Especially, the
contribution to entropy from superradiant modes is carefully
incorporated, leading to a result for this contribution which corrects
some previous errors in the literature. It turns out that the previous
errors were due to an incorrect quantization of the superradiant modes.
Some of main results for the case of rotating BTZ black holes are that
the entropy contribution from superradiant
modes is positive rather than negative and also has a leading order
divergence as that from nonsuperradiant modes. The total entropy,
however, can still be identified with the Bekenstein-Hawking entropy
of the rotating black hole by introducing a universal brick-wall cutoff.
Our correct treatment of superradiant modes in the ``angular-momentum
modified canonical ensemble'' also removes unnecessary introductions
of regulating cutoff numbers as well as ill-defined expressions in the
literature.

\vspace{1cm}
\noindent
%
\end{abstract}
\end{titlepage}

\newpage

Since Bekenstein \cite{Beken} suggested that black holes carry an
intrinsic entropy proportional to the surface area of the event
horizon, and Hawking \cite{Haw} provided a physical basis for this
idea by considering quantum effect, there have been various
approaches to understanding the black hole entropy. One of them is
the so-called ``brick-wall model'' introduced by 't Hooft
\cite{Hoof}. He has considered a quantum gas of scalar particles
propagating just outside the event horizon of the Schwarzschild
black hole. The entropy obtained just by applying the usual
statistical mechanical method to this system turns out to be
divergent due to the infinite blue shift of waves at the horizon.
't Hooft, however, has shown that the leading order term on the
entropy has the same form as the Bekenstein-Hawking formula for
the black hole entropy by introducing a brick-wall cutoff which is
a property of the horizon only and is the order of the Planck
length. The appearance of this divergence~\cite{SUT,DLM} and
relationships of this ``statistical-mechanical" entropy of quantum
fields near a black hole with its entanglement entropy~\cite{SE}
and quantum excitations of the black hole~\cite{BFZ} have been
studied, leading a great deal of interest recently~\cite{FF}.

The brick-wall model originally applied to the four dimensional
Schwarzschild black hole \cite{Hoof} has been extended to various
situations.
The application to the case of rotating black holes has also been done
for scalar fields in BTZ black holes in three-dimensions~\cite{KKPS,IS}
and in Kerr-Newman and other rotating black holes in four-dimensions
\cite{LK,HKPS}. In a background spacetime of rotating black holes, it
is well known that scalar fields have a special class of mode
solutions, giving superradiance. It is claimed in Ref.~\cite{IS} that
the statistical-mechanical entropy of a scalar matter is not
proportional to the ``area''({\it i.e.}, the circumference in the
three-dimensional case) of the horizon of a rotating BTZ black
hole and that the divergent parts are not necessarily due to the
existence of the horizon. Contrary to it, in Ref.~\cite{KKPS}, the
leading divergent term on the entropy is proportional to the ``area''
of the horizon, and it is possible to introduce a universal brick-wall
cutoff which makes the entropy equivalent to the black hole
entropy. Moreover, it is claimed in Ref.~\cite{KKPS} that the contribution from
superradiant modes to entropy is {\it negative} and its divergence is
in a {\it subleading} order compared to that from nonsuperradiant
modes. On the other hand, for the case of Kerr black holes in
Ref.~\cite{HKPS}, the divergence is in the leading order but the
entropy contribution is still negative.

One may expect that the leading contribution to the entropy comes from the region
very near the horizon as in the case of nonrotating black holes.
Since the vicinity of a rotating horizon can also be approximated by
the Rindler metric, it is seemingly that the essential feature of the leading
contribution will be same as that in nonrotating cases.
Our study in detail shows this naive expectation is indeed true. That is, we point out
that previous erroneous results appeared in the literature were
mainly related to an incorrect quantization of superradiant modes.
For the case of rotating BTZ black holes, we have explicitly shown that
superradiant modes also give leading order divergence to the entropy
as nonsuperradiant ones. Moreover, its entropy contribution is
{\it positive} rather than negative found in Refs.~\cite{KKPS,HKPS}.
However, the total entropy of quantum field can still be identified
with the Bekenstein-Hawking entropy by
introducing a universal brick-wall cutoff. It also has been shown that
the correct quantization of superradiant modes in the calculation of the
``angular-momentum modified canonical ensemble'' removes various
unnecessary regulating cutoff numbers as well as ill-defined
expressions in the literature.


\vspace{0.5cm}
Let us consider a quantum gas of scalar particles confined in a box near the
horizon of a stationary rotating black hole. The free scalar field satisfies the
Klein-Gordon equation given by $(\Box + \mu^2)\phi = 0$ with  periodic boundary conditions
\beq
\phi (r_++h) = \phi (L).
\label{BC}
\eeq
Here, $r_+$, $r_++h$, and $L$ are the radial coordinates of the horizon
and the inner and outer walls of a ``spherical" box, respectively.
Suppose that this boson gas is in a thermal equilibrium state at
temperature $\beta^{-1}$. Due to the existence of 
an ergoregion just outside the event horizon, any thermal
system sitting in this region must rotate with respect to an observer
at infinity. Accordingly, in order to obtain the appropriate grand
canonical ensemble for this rotating thermal system, one should
introduce an angular momentum reservoir as well in addition to a heat
bath/particle reservoir characterized by temperature $T = \beta^{-1}$
and angular speed $\Om$ with respect to an observer at infinity
\cite{LPW}. All thermodynamic quantities can be derived by the
partition function $Z(\beta , \Om )$ given by
\beq
Z(\beta , \Om ) = {\rm Tr} \, e^{-\beta :(\hat{H}- {\scriptsize \Om}
  \hat{J}):},
\label{Z0}
\eeq
where $:\!\!\hat{H}\!\!:$ and $:\!\!\hat{J}\!\!:$ are the normal ordered
Hamiltonian and angular momentum operators of the quantized field,
respectively~\cite{VA,LPW,FF}. Here, we assume that particle number of 
the system is indefinite. 

As usual, by using the single-particle spectrum,
one can obtain the free energy $F(\beta , \Om )$ of the system
in the following form
\beqa
\beta F &=& - \ln Z = -\sum_{\la} \ln \sum_{k} [e^{-\beta (\ve_{\la}
  -\so j_{\la})}]^{k}
\label{F0}    \\
&=& \left\{ \begin{array}{cc}
    -\sum_{\la}\ln [1+e^{-\beta (\ve_{\la}-\so j_{\la})}] &
    \mbox{for fermions,}  \\
    \sum_{\la} \ln [1-e^{-\beta (\ve_{\la}-\so j_{\la})}] &
    \mbox{for bosons with $\ve_{\la}-\Om j_{\la} > 0$.}
\end{array}
\right.
\eeqa
where $\la$ denotes the single-particle states for the free gas in the
system. $\ve_{\la}= \langle 1_{\la} |:\!\!\hat{H}\!\!:|1_{\la} \rangle $
and $j_{\la}= \langle 1_{\la} |:\!\!\hat{J}\!\!:|1_{\la} \rangle $ are normal ordered
energy and angular momentum associated with single-particle states $\la$,
respectively. The occupation number $k = 0, 1, 2, \cdots $ for bosonic
fields and $k=0, 1$ for fermionic fields.
Note that, if there exists a bosonic single-particle state with its
energy $\ve_{\la}$ and angular momentum $j_{\la}$ such that
$\ve_{\la}-\Om j_{\la} < 0$, the expression in Eq.~(\ref{F0}) becomes
divergent and so is ill-defined. In order to compute the free energy
in Eq.~(\ref{F0}), one must know all single-particle states and their
corresponding values of $\ve_{\la}$ and $j_{\la}$ for a given system.

As pointed out in Refs.~\cite{FT,MDOF}, the quantization of matter fields on
a stationary {\it rotating} axisymmetric black hole background is somewhat
unusual due to superradiant modes which occur in the presence of an ergoregion.
The mode solutions will be of the form, 
$\phi (x) \sim e^{-i{\scriptsize \om }t+im\varphi }$,
because this background spacetime possesses two Killing vector fields denoted by
$\pa_t$ and $\pa_{\varphi}$.
Since the partition function in Eq.~(\ref{Z0}) is defined with respect to an observer
at infinity, the vacuum state to be defined by the standard quantization procedure
should be natural to that observer at infinity in the far future. Thus, we expand
the neutral scalar field in terms of a complete set of mode solutions as follows:
\beqa
\phi (x) &=& \sum_m\int^{\infty}_{0}d\om (b^{\rm out}_{\scriptsize
             \om m}u^{\rm out}_{\scriptsize \om m} + b^{\dagger
             {\rm out}}_{\scriptsize \om m}u^{\ast {\rm out}}_{
             \scriptsize \om m}) + \sum_m\int^{\infty}_{\scriptsize
             m\Om_{H}}d\om (b^{\rm in}_{\scriptsize \om m}u^{\rm in}_{
             \scriptsize \om m} + b^{\dagger {\rm in}}_{\scriptsize
             \om m}u^{\ast {\rm in}}_{\scriptsize \om m})   \\
& & + \sum_m\int_{0}^{\scriptsize m\Om_{H}}d\om (b^{\rm in}_{
    \scriptsize -\om -m}u^{\rm in}_{\scriptsize -\om -m}
    + b^{\dagger {\rm in}}_{\scriptsize -\om -m}
    u^{\ast {\rm in}}_{\scriptsize -\om -m}).
\eeqa
Here $u^{\rm out}(x)$ describes unit outgoing flux to the future null infinity
${\cal T}^+$ and zero ingoing flux to the horizon ${\cal H}^+$
while $u^{\rm in}(x)$ describes unit ingoing flux to ${\cal H}^+$ and zero
outgoing flux to ${\cal T}^+$. These mode solutions are orthonormal
\beq
\langle u^{\rm out}_{\scriptsize \om m}\, ,\,
u^{\rm out}_{\scriptsize \om'm'} \rangle
= \langle u^{\rm in}_{\scriptsize \om m}\, ,\,
u^{\rm in}_{\scriptsize \om'm'} \rangle
= \langle u^{\rm in}_{\scriptsize -\om -m}\, ,\,
u^{\rm in}_{\scriptsize -\om' -m'} \rangle
= \delta (\om -\om')\delta_{mm'}
\eeq
with respect to the Klein-Gordon inner product
\beq
\langle \phi_1\, ,\, \phi_2 \rangle
 = \frac{i}{2}\int_{t={\rm const.}} \phi_1^{\ast}\!
   \stackrel{\leftrightarrow }{\partial_{\mu}}\! \phi_2\,\,
   d\Sigma^{\mu}.
\eeq Note that $u_{\scriptsize \om m}(x) \sim e^{\scriptsize -i\om
t+im\varphi}$ and we suppressed other quantum numbers. Modes with
$\tilde{\om}=\om -\Om_H m < 0$ exhibit the so-called
superradiance. Here $\Om_H$ is the angular speed of the horizon
with respect to an observer at infinity. An observer at infinity
would measure positive frequency for all modes $u^{\rm
out}_{\scriptsize \om m}$ and $u^{\rm in}_{\scriptsize \om m}$
with $\tilde{\om} >0$, but measure negative frequency for $u^{\rm
in}_{\scriptsize -\om -m}$ with $\tilde{\om} <0$. A
ZAMO~\cite{ZAMO} near the horizon, however, would see positive
frequency waves for $u^{\rm in}_{\scriptsize -\om -m}$ with
$\tilde{\om} <0$ as well as for $u^{\rm in}_{\scriptsize \om m}$
with $\tilde{\om} >0$. Hence, in the terminology of
Ref.~\cite{FT}, we adopt the ``distant-observer viewpoint" for
$u^{\rm out}(x)$ and the ``near-horizon viewpoint" for $u^{\rm
in}(x)$. And the conventions are chosen so that they agree with
viewpoints.

Now the mode solutions for particles confined in the near-horizon box
would be constructed by linearly superpose $u^{\rm in}$ and $u^{\rm out}$ above
as follows:
\beq
\phi_{\scriptsize \om m}(x) \sim \left\{ \begin{array}{cc}
     u^{\rm out}_{\scriptsize \om m} + \alpha_{\scriptsize \om m}
     u^{\rm in}_{\scriptsize \om m} &
     \mbox{for $\tilde{\om} >0$,}  \\
     u^{\rm out}_{\scriptsize \om m} + \alpha_{\scriptsize \om m}
     u^{\ast \rm in}_{\scriptsize -\om -m} &
     \mbox{for $\tilde{\om} <0$,}
\end{array}
\right.
\eeq
with appropriate normalization factor. $\alpha_{\scriptsize \om m}$ is chosen so that
the modes satisfy the periodic boundary condition in Eq.~(\ref{BC}).
Thus, only some discrete (real) values of $\om$ will be allowed~\cite{comp}.
$\phi_{\scriptsize \om m}(x)$ are understood to be cut off everywhere outside the box.
Note that $\phi_{\scriptsize \om m}(x) \sim e^{\scriptsize -i\om t+im\varphi}$
for all $\tilde{\om}$.

The inner product of these modes becomes
\beq
\langle \phi_{\scriptsize \om m}\, ,\, \phi_{\scriptsize \om'm'} \rangle
 = \de_{\scriptsize \om \om'}\de_{\scriptsize mm'}\!\! \int^{L}_{r_++h}
 (\om -\0 m)|\phi_{\scriptsize \om m}|^2 N^{-1}d\Sigma ,
\label{Inner}
\eeq
where we have used $d\Sigma^{\mu}=n^{\mu}d\Sigma$ and the unit normal
to a $t= {\rm const.}$ surface
$n^{\mu}=N^{-1}(\pa_t+\Om_0\pa_{\varphi})^{\mu}$. Here $\Om_0(r)$ is the angular
speed of ZAMO's~\cite{ZAMO}. Since $\Om_0(r) \leq \Om_H=\Om_0(r=r_+)$,
the norm of a mode solution with $\om > 0$
is positive if $\tilde{\om}=\om -\Om_Hm > 0$.
When $\tilde{\om} < 0$, the norm could be either positive or negative
depending on the radial behavior of the solution. If the norm of
$\phi_{\scriptsize \om m}(x)$ is negative,
we can easily see that $\phi_{\scriptsize -\om -m}(x) \sim
e^{\scriptsize i\om t-im\varphi}$
has the positive norm. Let us define a set SR consisting of
mode solutions $\phi_{\scriptsize \om m}$ with $\om >0$ whose norms are
negative. Then, the quantized field inside the box can
be expanded in terms of orthonormal mode solutions as
\beq
\phi (x) = \sum_{\la \not\in {\rm SR}}[a_{\scriptsize \om m}
  \phi_{\scriptsize \om m}(x) +
  a^{\dagger}_{\scriptsize \om m}\phi^{\ast}_{\scriptsize \om m}(x)]
  + \sum_{\la \in {\rm SR}}[a_{\scriptsize -\om -m}
  \phi_{\scriptsize -\om -m}(x) + a^{\dagger}_{\scriptsize -\om -m}
  \phi^{\ast}_{\scriptsize -\om -m}(x)],
\label{field}
\eeq
where the single-particle states are labeled by
$\la =(\om ,m)$. The Hamiltonian operator in the reference frame
of a distant observer at infinity becomes then \beqa H &=&
\sum_{\la \not\in {\rm SR}} \om (a_{\scriptsize \om m}a^{\dagger}_
      {\scriptsize \om m} + a^{\dagger}_{\scriptsize \om m}
      a_{\scriptsize \om m}) + \sum_{\la \in {\rm SR}}(-\om )
      (a_{\scriptsize -\om -m}a^{\dagger}_{\scriptsize -\om -m}
      + a^{\dagger}_{\scriptsize -\om -m}a_{\scriptsize -\om -m})
      \nonumber   \\
&=& \sum_{\la \not\in {\rm SR}} \om (N_{\scriptsize \om m}+\frac{1}{2})
    +\sum_{\la \in {\rm SR}}(-\om )(N_{\scriptsize -\om -m}+\frac{1}{2}),
\eeqa where $N_{\scriptsize \om m}=a^{\dagger}_{\scriptsize \om m}
a_{\scriptsize \om m}$ and $N_{\scriptsize -\om -m}=a^{\dagger}_
{\scriptsize -\om -m}a_{\scriptsize -\om -m}$ are number
operators. Now, by following the standard procedure for defining a
vacuum state and single-particle states~\cite{MDOF,FT}, we can
easily see that $(\ve_{\la},j_{\la})=(\om ,m)$ for single-particle
states $\la =(\om ,m) \not\in {\rm SR}$ while
$(\ve_{\la},j_{\la})=(-\om ,-m)$, instead of $(\om ,m)$, for
single-particle states $\la = (\om ,m) \in {\rm SR}$. In
Refs.~\cite{KKPS,IS,HKPS}, however, $(\ve_{\la},j_{\la})=(\om ,m)$
for $\la \in {\rm SR}$ have been used, and our study shows that
this error comes from the incorrect quantization of superradiant
modes. This important difference is a peculiar feature of the
quantization of matter fields in the presence of an ergoregion and
turns out to make our ``angular-momentum modified canonical
ensemble'' in Eq.~(\ref{Z0}) being well-defined as shall be shown
below in detail. It also makes somewhat unphysical treatment of
superradiant modes and introduction of various cutoff numbers
unnecessary in the calculation of statistical-mechanical entropy
in the literature. For example, in Ref.~\cite{IS}, a cutoff in the
occupation number $k$ was introduced to avoid the divergent sum
for superradiant modes in the log in Eq.~(\ref{F0}).

In general, $\om$ is discrete due to the finite size of the box,
but the gap between adjacent values
goes small as the size of the thermal box becomes large.
In this continuous limit, one may introduce the density function
defined by $g(\om ,m)=\pa n(\om ,m)/\pa \om$ where $n(\om ,m)$ is
the number of mode solutions whose frequency or energy is below $\om$
for a given value of angular momentum $m$. Thus, $g(\om ,m)d\om$
represents the number of single-particle states whose energy lies
between $\om$ and $\om +d\om$, and whose angular momentum is $m$.
Using this density function, the free energy in Eq.~(\ref{F0}) can be
re-expressed as
\beq
\beta F= -\sum_m \!\!\int \!d\om \, g(\om ,m) \ln \sum_{k}
         [e^{-\beta (\ve_{\la} -\so j_{\la})}]^{k}.
\label{F1}
\eeq

The angular speed $\Om$ in Eq.~(\ref{F1}) is a thermodynamic parameter
defined, in principle, by its appearance in the thermodynamic first
law for the reservoir, namely $TdS=dE-\Om dJ+ \cdots $. Since a particle
cannot move faster than the speed of light, its angular velocity with respect
to an observer at rest at infinity should be restricted. The possible maximum and minimum
angular speeds are
\beq
\Om_{\pm}(r) = \Om_0 (r) \pm \sq{(\pa_t \cdot \pa_{\varphi}/\pa_{\varphi}
\cdot \pa_{\varphi})^2 - \pa_t \cdot \pa_t/\pa_{\varphi}\cdot \pa_{\varphi}},
\label{Ompm}
\eeq
respectively. We see that, as $r \rightarrow r_+$, the range of
angular velocities a particle can take on narrows down({\it i.e.},
$\Om_{\pm}(r) \rightarrow \Om_H$), and so the angular speed of particles near
the horizon will be $\Om_H$. For a rotating body in flat spacetime, one knows
that all subsystems must rotate uniformly when the body is
in a thermal equilibrium state \cite{LPW}. In fact, it is a part of
the thermodynamic zeroth law.
In curved spacetimes, the uniform rotation of all
subsystems in thermal equilibrium may not be true to hold any more.
However, since we will be finally interested in the quantum gas only in the
vicinity of the horizon, we shall assume $\Om =\Om_H$ below.

Now, one can see that the sum in the log in Eq.~(\ref{F1}) is defined
well for states belonging to SR since 
$\ve_{\la}-\Om j_{\la} = -(\om -\Om_H m) >0$ by the definition of the set SR. 
For some states with $\om -\Om_Hm < 0$ 
not belonging to SR, however, the sum becomes
divergent. As will be shown below explicitly, however, the main
contribution to the entropy of the system comes from the infinite
piling up of waves at the horizon. For such localized solutions near
the horizon, the signature of norms in Eq.~(\ref{Inner}) will be
determined by that of $\om -\Om_Hm$ since $\Om_0(r) \simeq \Om_H$ for
the range of integration giving dominant contributions. Therefore, we
assume that all states with $\om -\Om_Hm < 0$ belong to the set SR.
Then, the free energy in Eq.~(\ref{F1}) can be written as
$F = F_{\rm NS} +F_{\rm SR}$.
Here
\beqa
\be F_{\rm NS} &=& \sum_{\la \not\in {\rm SR}}\!\!\int d\om \, g(\om ,m)
\ln [1-e^{-\beta (\om -\somh m)}],
\label{FNS}           \\
\be F_{\rm SR} &=& \sum_{\la \in {\rm SR}}\!\!\int d\om \, g(\om ,m)
\ln [1-e^{\beta (\om -\somh m)}].
\label{FSR}
\eeqa

In general cases, it is highly nontrivial to compute $g(\om ,m)$
exactly except for some cases in two-dimensional black
holes~\cite{SCZ}. For suitable conditions, however, one can
approximately obtain $g(\om ,m)$ by using the WKB method as in the
brick-wall model~\cite{Hoof}. For simplicity, let us consider a
scalar field in a rotating BTZ black hole in
3-dimensions~\cite{BTZ}. For the case of Kerr black holes in
4-dimensions, although the essential result is the same as that in
the case of BTZ black holes, it requires some modified formulation
basically due to the fact that the geometrical property near the
horizon changes along the polar angle~\cite{Kerr}. The metric of a
rotating BTZ black hole is given by \beq ds^2= -N^2dt^2 +
N^{-2}dr^2 + r^2(d\varphi - \0 dt)^2, \eeq where \beq N^2= r^2/l^2
-M +J^2/4r^2 = (r^2-r^2_+)(r^2-r^2_-)/r^2l^2,  \nonumber \eeq and
the angular speed of ZAMO's is $\Om_0 =J/2r^2$. Here $r_{\pm}$
denote the outer and inner horizons, respectively. Note that
$\Om_H=\Om_0(r=r_+)=r_-/r_+l $. Then, mode solutions are
$\phi_{\scriptsize \om m}(x) = \phi_{\scriptsize \om m}(r)
e^{-i{\scriptsize \om}t+im\varphi }$. Here the radial part
$\phi_{\scriptsize \om m}(r)$ satisfies \beq rN^2
\frac{d}{dr}\Big[rN^2\frac{d}{dr}\phi_{\scriptsize \om m}(r)\Big]
+ r^2N^4k^2(r; \om ,m)\phi_{\scriptsize \om m}(r) =0, \label{REQ}
\eeq where \beq k^2(r;\om ,m) = N^{-4}[(\om -\+ m)(\om -\-
m)-\mu^2N^2] . \eeq Here $\Om_{\pm}(r) = \Om_0 (r) \pm N/r$ for a
rotating BTZ black hole in Eq~(\ref{Ompm}).

In the WKB approximation, the discrete value of energy $\om$
in Eq.~(\ref{REQ}) is related to $n(\om ,m)$ as follows
\beq
\pi n(\om ,m) = \int^{L}_{r_++h} dr\, ``k\textrm{''}(r;\om ,m),
\label{WKB}
\eeq
where $``k\textrm{''}(r;\om ,m)$ is set to be zero if $k^2(r;\om ,m)$
becomes negative for given $(\om ,m)$ \cite{Hoof}.
Since $``k\textrm{''}(r;\om ,m) \simeq N^{-2}$
and $N(r) \rightarrow 0$ as one approaches to the horizon,
we can easily see that the dominant
contribution in Eq.~(\ref{WKB}) comes from the integration in the
vicinity of the horizon as the inner brick-wall approaches to it(
{\it i.e.}, $h \rightarrow 0$).

Now, Eq.~(\ref{FNS}) becomes
\beqa
\beta F_{\rm NS} &=& \sum_{\la \not\in {\rm SR}}\!\int \!\! d\om \,
  \f{\pa}{\pa \om}\bigg[ \f{1}{\pi}\int^L_{r_++h}\!\! dr \,
  ``k\textrm{''}(r;\om ,m)\bigg] \ln [1-e^{-\beta (\om -\somh m)}]
  \nonumber   \\
&=& -\frac{\be}{\pi} \int^L_{r_++h}\!\! dr \sum_{m}\!\int \!\! d\om
  \, \frac{k(r;\om ,m)}{e^{\beta (\om -\somh m)}-1}   \nonumber   \\
& & +\frac{1}{\pi} \int^L_{r_++h}\!\! dr \sum_{m}\, k(r;\om ,m)
  \ln [1-e^{-\beta (\om -\somh m)}] \Big|^{\om_{\rm max}(m)}_{\om_{\rm
      min}(m)}
\label{F2}
\eeqa
by using the integration by parts in $\om$.
For convenience, one can divide $F_{\rm NS}$ into two parts
\beq
F_{\rm NS} = F^{(m>0)}_{\rm NS} + F^{(m<0)}_{\rm NS}, \nonumber
\eeq
where
\beq
F^{(m>0)}_{\rm NS} =  -\frac{1}{\pi}\int^L_{r_++h}\!\! dr \, N^{-2}
\int^{\infty}_0 \!\! dm \!\!\int^{\infty}_{\+ m}\!\! d\om \,
\frac{\sq{(\om -{\scriptsize \Om_+} m)(\om -{\scriptsize \Om_-}
    m)}}
{e^{\beta (\om -\somh m)}-1}
\label{NS1+}
\eeq
from states with positive angular momenta and
\beqa
F^{(m<0)}_{\rm NS} &=& -\f{1}{\pi}\bigg( \int^{r_{\rm erg}}_{r_++h}
  dr\, N^{-2}\int^{0}_{-\infty}dm \int^{\infty}_{0}d\om  \nonumber \\
& & +\int^{L}_{r_{\rm erg}}dr\, N^{-2}\int^{0}_{-\infty}dm
  \int^{\infty}_{\- m}d\om \bigg)\, \f{\sq{(\om -{\scriptsize \Om_+} m)
  (\om -{\scriptsize \Om_-} m)}}{e^{\beta (\om -\somh m)}-1}
  \nonumber  \\
& & -\f{1}{\pi \be}\int^{r_{\rm erg}}_{r_++h}dr\, N^{-2}
  \int^{0}_{-\infty}dm\, \sq{\+ \- m^2} \ln (1-e^{\beta \somh m})
\label{NS1-}
\eeqa
from states with negative angular momenta. $r_{\rm erg}=\sq{M}l$ 
is the radius of the outer boundary of the ergoregion where 
$\Om_-(r=r_{\rm erg})=0$. Here we considered a
massless scalar field for simplicity.
Similarly, from states belonging to SR, Eq.~(\ref{FSR}) becomes
\beqa
F_{\rm SR} &=& -\f{1}{\pi}\int^{r_{\rm erg}}_{r_++h}dr\, N^{-2}
  \int^{\infty}_{0}dm \int^{\- m}_{0} d\om \, \f{\sq{(\om
      -{\scriptsize \Om_+} m)(\om -{\scriptsize \Om_-} m)}}
  {e^{-\beta (\om -\somh m)}-1}  \nonumber  \\
& & +\f{1}{\pi \be}\int^{r_{\rm erg}}_{r_++h}dr\, N^{-2}
  \int^{\infty}_{0}dm \sq{\+ \- m^2}\, \ln (1-e^{-\beta \somh m}).
\label{SR}
\eeqa
Note that $g=-\pa n/\pa \om$ for $\la \in {\rm SR}$, and 
that the boundary term in $F_{\rm SR}$ exactly cancels that in
$F^{(m<0)}_{\rm NS}$.

{}From Eqs.~(\ref{NS1+}-\ref{SR}), we can obtain leading order
dependence on the brick-wall cutoff $h$ for the free energy as
follows; \beqa F^{(m>0)}_{\rm NS} &=& -\f{\zeta (3)}{\be^3}
\f{r^2_{+}l^3}
  {(r^2_{+}-r^2_{-})^2} \left[ \f{\sq{r^2_+-r^2_-}}{2\sq{2}}
  \sq{\f{r_+}{h}} + \f{r_-}{\pi}\ln (\f{r_+}{h}) +
  \vartheta (\sq{h}) \right] ,    \nonumber  \\
F^{(m<0)}_{\rm NS} &=& -\f{\zeta (3)}{\be^3} \f{r^2_{+}l^3}
  {(r^2_{+}-r^2_{-})^2} \left[ \f{r^2_+-r^2_-}{2\pi r_-} \ln (\f{r_+}{h})
  + \vartheta (\sq{h}) \right] ,    \nonumber   \\
F_{\rm SR} &=& -\f{\zeta (3)}{\be^3} \f{r^2_{+}l^3}
  {(r^2_{+}-r^2_{-})^2} \Bigg[ \f{\sq{r^2_+-r^2_-}}{2\sq{2}}
  \sq{\f{r_+}{h}} - \f{r_-}{\pi}\ln (\f{r_+}{h})
  - \f{r^2_+-r^2_-}{2\pi r_-} \ln (\f{r_+}{h})  \nonumber  \\
& & + \vartheta (\sq{h}) \Bigg] .
\label{Free}
\eeqa
The entropy of this boson gas which is assumed to be in thermal
equilibrium with the rotating black hole can be obtained from the free
energy by using the thermodynamic relation,
$S = \be^2 \pa F/\pa \be |_{\scriptsize \be = \be_H}
=-3\beta F|_{\scriptsize \be = \be_H}$;
\beqa
S_{\rm NS} &=& \f{3\zeta (3)}{4\pi^2l}
  \left[ \f{\sq{r^2_+-r^2_-}}{2\sq{2}}
  \sq{\f{r_+}{h}} + \f{r_-}{\pi}\ln (\f{r_+}{h}) +
  \f{r^2_+-r^2_-}{2\pi r_-} \ln (\f{r_+}{h}) +
  \vartheta (\sq{h}) \right] ,   \nonumber  \\
S_{\rm SR} &=& \f{3\zeta (3)}{4\pi^2l}
  \left[ \f{\sq{r^2_+-r^2_-}}{2\sq{2}}
  \sq{\f{r_+}{h}} - \f{r_-}{\pi}\ln (\f{r_+}{h})
  - \f{r^2_+-r^2_-}{2\pi r_-} \ln (\f{r_+}{h})
  + \vartheta (\sq{h}) \right] ,
\label{ENTS}
\eeqa
where the temperature of a rotating BTZ black hole~\cite{BTZ} is
\beq
\be^{-1}_{H} = (r^2_+ -r^2_-)/2\pi r_+l^2.
\eeq
Now the total entropy of the system becomes
\beq
S = \f{3\zeta (3)}{4\sq{2}\pi^2} \f{\sq{r^2_+-r^2_-}}{l}
    \sq{\f{r_+}{h}} + \vartheta (\sq{h}).
\label{ENT}
\eeq

In Ref.~\cite{KKPS}, it is claimed that the contribution from
superradiant modes is a subleading order compared with that from
nonsuperradiant modes. In our results above, however, we find that 
superradiant modes also give a leading order contribution which is in fact 
exactly same as that from nonsuperradiant modes in the leading order of
$\sqrt{r_+/h}$.
It should be pointed out that
the entropy associated with superradiant modes is {\it positive} in
our result whereas it is {\it negative} in Refs.~\cite{KKPS,HKPS}.
In addition, since the log terms in Eq.~(\ref{ENTS}) are exactly cancelled,
our result for the entropy of quantum field smoothly
reproduces the correct result in the non-rotating limit({\it i.e.,}
$J \rightarrow 0$ or $r_- \rightarrow 0$) whereas the entropy obtained in
Refs.~\cite{KKPS,HKPS} becomes divergent in that limit.

If we rewrite the entropy in terms of the brick-wall cutoff in
proper length defined as $\bar{h} = \int^{r_++h}_{r_+} \sq{g_{\rm
rr}}dr$, Eq.~(\ref{ENT}) becomes \beq S = \f{3\zeta
(3)/8\pi^3}{\bar{h}} {\cal C} + \vartheta (\bar{h}), \eeq where
${\cal C}=2\pi r_+$ is the circumference of the horizon. Thus, by
recovering the dimension and introducing an appropriate brick-wall
cutoff \beq \bar{h} = \f{3\zeta (3)}{16\pi^3} l_P \simeq 7.3
\times 10^{-3} l_P \label{CUT} \eeq which is a universal constant,
one can make the entropy of quantum field finite and being
equivalent to the Bekenstein-Hawking entropy of a rotating BTZ
black hole~\cite{BTZ} \beq S = 4\pi r_+/l_P = S_{\rm BH} \eeq in
leading order. Here $l_P$ is the Planck length. For a fermionic
field, although modes with $\tilde{\om} <0$ do not reveal
superradiance, it turns out that only the overall numerical factor
in Eq.~(\ref{ENT}) is different. As mentioned before, the
extension of our study to the case of Kerr black holes in
four-dimensions is straightforward, but requires some
modifications mainly due to the polar angle dependence of the near
horizon geometry. A calculation in the phase space shows that the
essential feature of the leading order divergence in the entropy
of quantum fields is same as that of the present case~\cite{Kerr}.

Other thermodynamic quantities of quantum field such as the angular
momentum and internal energy can also be obtained as follows;
\beq
J_{\rm matter} = -\f{\pa F}{\pa \Om}{\bigg|}_{\scriptsize
  \be =\be_H, \Om =\Om_H}
  = \f{3\zeta (3)/16\pi^3}{\bar{h}}\, \f{2r_+r_-}{l}
   + \vartheta (\ln \bar{h}).
\eeq
Here the derivative with respect to $\Om$ has been taken for
Eqs.~(\ref{NS1+}-\ref{SR}).
If we put the cutoff value in Eq.~(\ref{CUT}), we have
\beq
J_{\rm matter} = \f{2r_+r_-}{l} = J_{\rm BH}.
\eeq
The internal energy of the system with respect to an observer at
infinity is
\beqa
E &=& \f{\pa}{\pa \be}(\be F){\bigg|}_{\scriptsize
  \be =\be_H, \Om =\Om_H} + \Om_HJ_{\rm matter}  \nonumber \\
  &=& \f{3\zeta (3)/16\pi^3}{\bar{h}}\,
  \f{4}{3}\f{r^2_++\f{1}{2}r^2_-}{l^2} +\vartheta (\ln \bar{h})
  = \f{4}{3}M_{\rm BH} - \f{2}{3} \f{r^2_-}{l^2},
\eeqa
where the black hole mass is $M_{\rm BH}=M=(r^2_++r^2_-)/l^2$.
One can easily see that $J_{\rm matter} \rightarrow 0$ and $E
\rightarrow \f{4}{3}M_{\rm BH}$ in the limit of non-rotating black
holes ({\it e.g.}, $J_{\rm BH}=J \rightarrow 0$). Therefore,
we find that the entropy and angular momentum of quantum field can be
identified with those of the rotating black hole by introducing
a universal brick-wall cutoff although the internal energy is
not proportional to the black hole mass.


\vspace{0.5cm}
What kind of relationships could be held among parameters characterizing
a rotating black hole and thermodynamic quantities of
the system of quantum fields
in equilibrium with the black hole? To see this, let us consider
a system whose free energy depends on the temperature such that
$F(\be ,\Om ,M,J) = \be^{-3}f(\Om ,M,J)$~\cite{footnote}.
Suppose the entropy of the system is identified with that
of the black hole after an appropriate regularization,
\beq
S = \be^2 \left(\f{\pa F}{\pa \be}\right)_{\scriptsize \Om}
{\bigg|}_{\scriptsize \be =\be_H, \Om =\Om_H} =
S_{\rm BH} = 4\pi r_+.
\eeq
The internal energy of the system with respect to a ``corotating''
observer is
\beq
E^{\prime} = \left[\f{\pa}{\pa \be}(\be F)\right]_{\scriptsize \Om}
{\bigg|}_{\scriptsize \be =\be_H, \Om =\Om_H}
           = \f{2}{3} \f{S}{\be_H} = \f{4}{3} \f{r^2_+-r^2_-}{l^2}.
\eeq
Now suppose the angular momentum of the system is proportional to that
of the rotating black hole,
\beq
J_{\rm matter} = -\left(\f{\pa F}{\pa \Om}\right)_{\scriptsize \be}
{\bigg|}_{\scriptsize \be =\be_H, \Om =\Om_H}
               = \al J = \al \f{2r_+r_-}{l}.
\eeq
The internal energy of the system with respect to an observer at
infinity
is then
\beq
E = E^{\prime} + \Om_H J_{\rm matter} = \f{4}{3} \f{r^2_+ + (3\al
  /2-1)r^2_-}{l^2},
\eeq
which becomes proportional to the mass of the black hole,
$M=(r^2_++r^2_-)/l^2$, only if $\al = 4/3$. Therefore, we expect the
relationships are probably
\beq
J_{\rm matter} = \f{4}{3} J, \qquad    E = \f{4}{3} M.
\label{EJ}
\eeq
If we apply the same argument to the case of Kerr black holes in
4-dimensions, we obtain
\beq
J_{\rm matter} = \f{3}{4} J,  \qquad  E = \f{3}{8} M
\eeq
of which the second relationship has been explicitly shown for the
Schwarzschild black hole in the brick-wall model 
by 't Hooft~\cite{Hoof}.

We have not obtained the relationships in Eq.~(\ref{EJ}) at the
present letter. The reason for these discrepancies is not understood
at the present. It will be very interesting to see how
the Pauli-Villars regularization method, which does not require
the presence of a brick-wall as shown in Ref.~\cite{DLM} for the case
of a charged non-rotating black hole, works for the case of
rotating black holes.

\vskip 0.7cm
Authors would like to thank, for useful discussions,
W.T. Kim, Y.J. Park, and H.J. Shin. Especially, GK would like to thank
J. Samuel, T. Jacobson, and H.C. Kim for many helpful discussions
and suggestions. GK was supported by Korea Research Foundation.



\begin{thebibliography}{99}
\bibitem
{ZAMO} ZAMO's are locally nonrotating observers whose trajectories
       are tangent to $\pa_{\tau}=\pa_t+\Om_0 \pa_{\varphi}$ where
       $\Om_{0}(r) =-\pa_t \cdot \pa_{\varphi}/{\pa_{\varphi}\cdot
       \pa_{\varphi}}$.
\bibitem
{footnote} One can show that in general this form of the free energy holds 
for a massless field by introducing new integration variables defined as 
$x=\beta \omega$ and $y=\beta m$ in Eq.~(\ref{F1}), or, more explicitly, in 
Eqs.~(\ref{NS1+}-\ref{SR}).
\end{thebibliography}
\end{document}